\documentclass[twocolumn,preprintnumbers,amsmath,amssymb]{revtex4}

\usepackage{amsmath}
\usepackage{amssymb}
\usepackage{epsfig}
\begin{document}

\title{Phase-Locked, Low-Noise, Frequency Agile Titanium:Sapphire Lasers for Simultaneous Atom Interferometers}
\author{Holger M\"uller, Sheng-wey Chiow, Quan Long, and Steven Chu.}
\affiliation{Department of Physics, Stanford University, Stanford, California 94305-4060}

\begin{abstract}
We demonstrate a laser system consisting of a $>1.6$\,W titanium:sapphire laser phase-locked to another free-running one at a wavelength of 852\,nm with a phase noise of \text{-138\,dBc/Hz} at 1\,MHz from the carrier, using an intra-cavity electro-optic phase modulator. The residual phase variance is $2.5\times10^{\text{-8}}$\,rad$^2$ integrated from 1\,Hz to 10\,kHz.  This system can change the offset frequency phase-continuously within 200\,ns with frequency steps up to 4\,MHz.  Simultaneous atom interferometers can make full use of this ultra-low phase noise in differential measurements, where influences from vibration of optics are greatly suppressed in common mode.
\end{abstract}

\maketitle 

\begin{figure}
\begin{center}
\epsfig{file=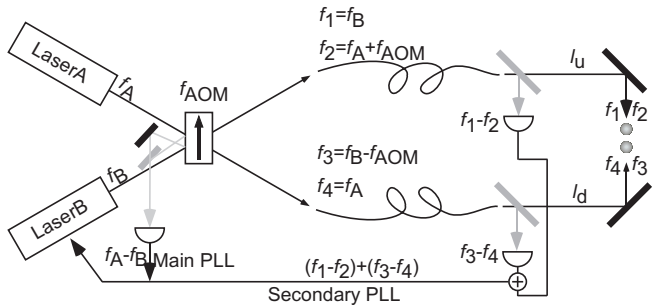,width=0.48\textwidth}
\caption{\label{conjugate}Frequency generation and phase locking scheme for simultaneous atom interferometers.}
\end{center}
\end{figure}

\begin{figure}
\begin{center}
\epsfig{file=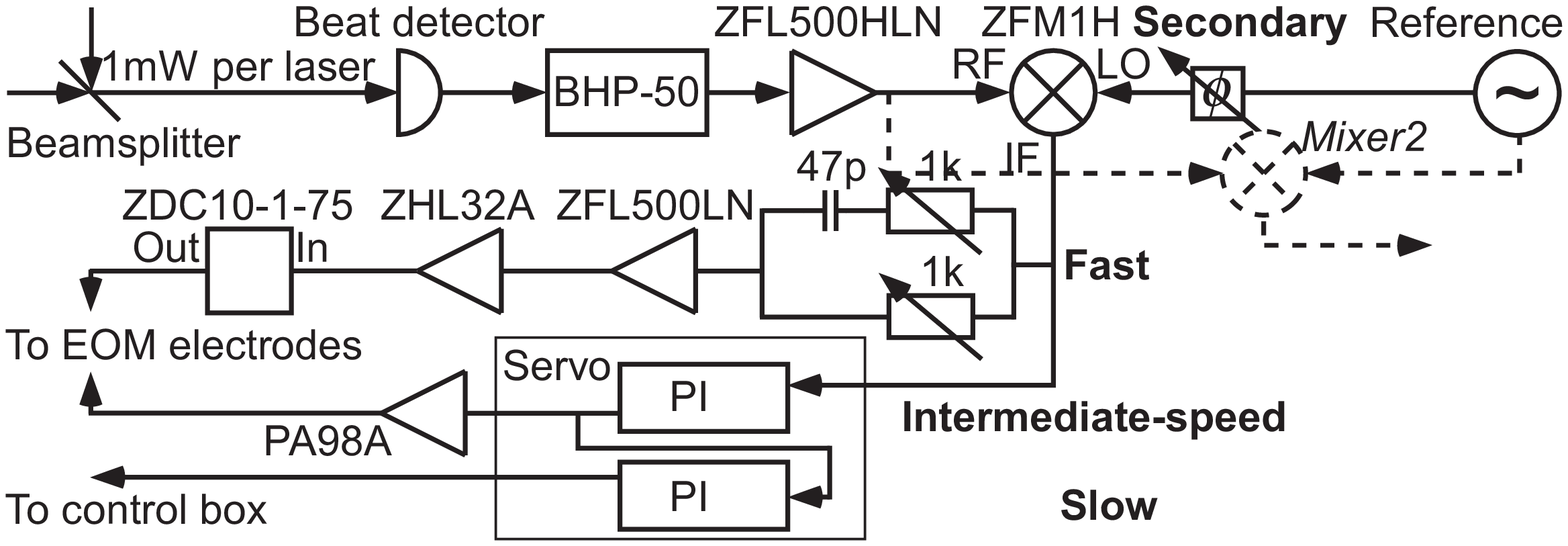,width=0.48\textwidth}
\caption{\label{schematic}Feedback loop block diagram.}
\end{center}
\end{figure}

\begin{figure*}[t]
\begin{center}
\epsfig{file=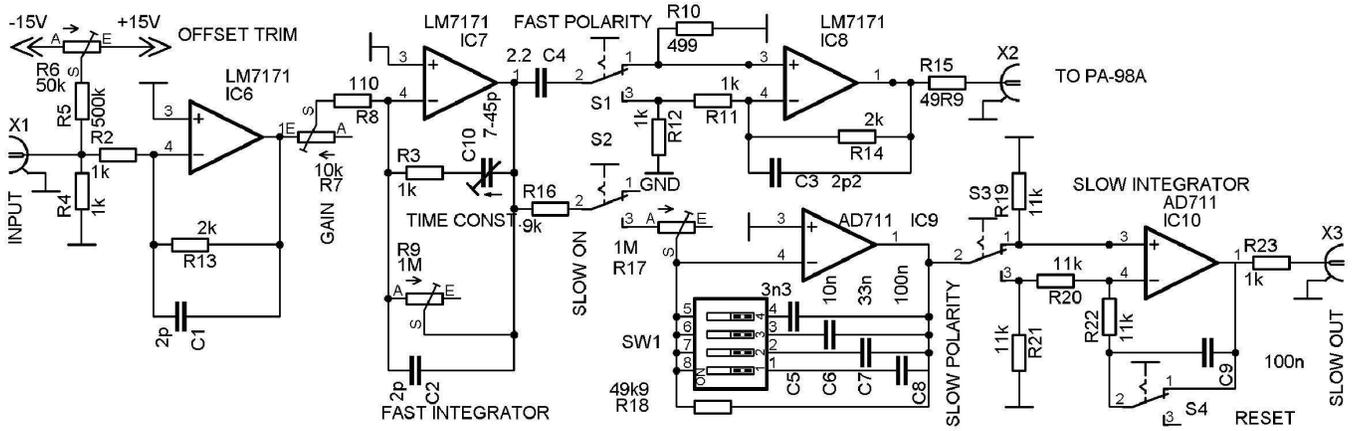,width=\textwidth}
\caption{\label{servo}Schematic of the slow and intermediate-speed PI controllers.}
\end{center}
\end{figure*}

\noindent The ability to precisely stabilize and manipulate the frequency and phase of laser light is at the basis of the tremendous progress in precision measurements, including the realization of optical clocks  \cite{opticalclock1,opticalclock2} and tests of fundamental laws of physics \cite{lorentz}.  This ability has been extended to phase-locking femtosecond lasers \cite{Shelton}.  Low-noise phase-locked continuous-wave (cw) lasers are also extensively used for cooling, condensing, and manipulating atoms \cite{natureinsight}; and in particular, atom interferometry is used in sensitive gyroscopes, precision gravimeters and gravity gradiometers, and precision measurements of fundamental constantsm \cite{berman}. In these applications, laser light serves as a reference for measuring the phase evolution of the matter waves.  Thus, lasers with ultra-low phase noise are required to achieve present atom interferometer accuracy goals.

To achieve low phase noise, fast feedback mechanisms are necessary to efficiently remove the frequency or phase fluctuations of the laser emission.  In such atom-manipulation systems, extended cavity diode lasers are widely used, because their output frequencies and phases can be easily controlled by changing the injection current.  Other than using fast feedback, low-noise phase-locking in diode laser systems can also be obtained by injection-locking \cite{injectionlocking}. For Nd:YAG lasers, which have an intrinsically small linewidth, the performance of cw phase-locked lasers can reach \text{-125\,dBc/Hz} at 10\,Hz offset from the carrier \cite{Ye}.  

In addition to ultra-low phase noise, applications in atomic physics often require high optical power and fast (microseconds), precise steps in frequency (`frequency-agility') \cite{contrastinter}.  
Although delicate phase-locked diode lasers can be frequency agile, these diode laser systems with optical power higher than 1\,W are rare, even with the help of commercially available optical power amplifiers. 
Haubrich and Wynands achieved phase-locking of a Ti:sapphire laser to a stable diode laser of 4\,kHz linewidth with a phase noise of about \text{-90\,dBc/Hz} at 1\,MHz offset by use of an intracavity electro-optic phase modulator (EOM) \cite{haubrich}. Here, we demonstrate a cw phase-locked system of two Ti:sapphire lasers, each providing more than 1.6\,W output power, with a phase noise of \text{-138\,dBc/Hz} at 1\,MHz from the carrier.  Our system can achieve phase-continuous frequency steps of 4\,MHz in 200\,ns.

Both master and slave lasers are Coherent 899-21 Ti:sapphire ring lasers in the high-power-pump configuration with control boxes modified as suggested by Haubrich and Wynands.  The master laser is pumped with 10.5\,W from a Coherent Verdi-10.  The frequency of the master laser is stabilized to its reference cavity with supplied servo elements and electronics, and the single-frequency output power is 1.6\,W at 852\,nm.  The master laser can be free-running, but will be referenced to a Cs vapor cell for frequency stability in our future applications.  The slave laser is pumped by a Coherent Innova \text{400-15} Argon ion laser with 19\,W multiline output power.  An EOM (Linos PM25IR) is inserted into the resonator between the Faraday rotator and the upper fold mirror.  This causes the output power to drop by $\sim20\%$ to 1.85\,W.  For locking, optical power is picked up from each Ti:sapphire laser using a residual reflection from a common optical element.  A total of 2\,mW reaches the beat detector (Fig.\ \ref{conjugate}).  The beatnote is amplified and a double-balanced mixer (DBM) serves as a phase detector, with the local oscillator (LO) input driven by an Agilent E8241A synthesizer at $\sim168$\,MHz.  In order to prevent laser amplitude fluctuations from coupling into the phase noise measurement, a 50\,MHz high-pass filter is placed before the phase detector.

The feedback to the slave laser consists of three paths: a slow, an intermediate-speed, and a fast path (Fig.\ \ref{schematic}).  The slow feedback path has a proportional-integral (PI) control with a bandwidth of 20\,kHz.  It controls the length of the laser's resonant cavity through the modified control box, which modulates a piezo-actuated mirror as well as etalons in the cavity.  
The fast and the intermediate-speed feedback paths feed to opposite electrodes of the intracavity EOM.  The intermediate-speed path uses a PI controller and a high-voltage operational amplifier (Apex PA98A, with a high slew rate of 1000\,V/\text{$\mu$s} and $\pm150$\,V swing) for driving the EOM, which provides a frequency tuning coefficient of 0.2\,MHz/V.  The bandwidth of this path is limited to 0.8\,MHz by PA98A.  The fast feedback path consists of a passive proportional-differential (PD) controller followed by an RF power amplifier (Fig.\ \ref{schematic}).  After this RF amplifier, a directional coupler is inserted before connecting to an electrode of the EOM to ground the electrode for DC voltages.  The differential control in the fast path compensates for phase lag due to propagation delay at high frequencies.  We achieve an overall closed-loop bandwidth of $\sim10$\,MHz, limited by the propagation delay in the optical and electronic signal path lengths.

\begin{figure}
\begin{center}
\epsfig{file=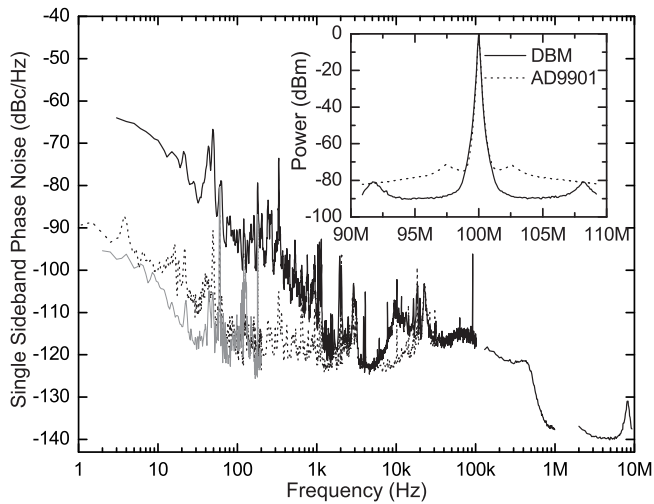,width=0.48\textwidth}
\caption{\label{noise}Residual phase noise with an analog phase detector (measured out of loop).  Solid line: phase lock before fiber only; dashed line: phase lock after fiber; gray line: electronic out-of-loop phase noise measured with Mixer2 in Fig.\ \ref{schematic}.  The gaps at 100\,kHz and 1\,MHz are due to different range settings between data sets.  Inset: phase noise spectra of analog and digital phase detectors.  Resolution bandwidth: 100\,kHz.}
\end{center}
\end{figure}

We study the performance of the phase-locked lasers using a separate beat detector after passing both beams through one single-mode polarization-maintaining fiber with the same polarization, for an out-of-loop measurement.  Figure\ \ref{noise} (solid line) shows the phase noise spectral density.  The wideband ($>100$\,kHz) noise, which goes down to \text{-138\,dBc/Hz}, is measured with an HP 8590B spectrum analyzer, and the low-frequency noise is measured with a SRS SR785 FFT signal analyzer by down-converting the beat signal to DC with a separate DBM but using the same LO as that in the PLL (This is found necessary, as otherwise the measurement is limited by the noise of LO.)  The peaks at 9\,MHz indicate the loop bandwidth.  The residual phase variance integrated over a frequency range from 1\,Hz to 10\,kHz is about $10^{\text{-6}}$\,rad$^2$ (solid line).

For comparison, we also measure the phase noise with a digital phase detector (Analog Devices AD9901) instead of a DBM (Fig.\ \ref{noise}).  The $\sim$8\,dB increase of the wideband phase noise probably arises from the conversion of the analog waveforms into digital signals, where voltage noise in the analog signal causes phase noise as it makes uncertain the time of zero-crossings.

The frequency agility is illustrated by stepping the LO frequency.  We find that with the DBM, phase-lock without cycle slips is maintained for sudden phase-continuous frequency changes as large as 4\,MHz.  This 4\,MHz frequency step is limited by the $\sim20$\,Vp dynamic range of the fast feedback path.  The transient behavior of the PLL is examined by adding a square wave signal to the output of the DBM.  The slave laser settles to 20\% within 200\,ns after each step.

The rise of phase noise at low frequencies (solid line in Fig.\ \ref{noise}) is mainly due to laser pointing fluctuations, air flow, and optics vibration.  This noise is not revealed in an electronically out-of-loop measurement (\text{-110\,dBc/Hz} at low frequencies, gray line in Fig.\ \ref{noise}) using the same photodetector but a separate mixer (Mixer2 in Fig.\ \ref{schematic}). (The best performance in this measurement is achieved by replacing the 17dBm types ZFM-1H by 27dBm types VAY-1, both by mini-circuits.)  Although this electronically out-of-loop performance is similar to the one reached by Ye and Hall \cite{Ye}, an atom interferometer still suffers the much higher noise.  

This is because the two beams must be sent to the atoms via separate paths in a counter-propagating geometry. Differential vibrational noise in these paths would essentially increase the noise to the level of the solid line in Fig.\ \ref{noise}, even if the noise is suppressed to a lower level at one point in the setup. 

We can avoid this and thus employ the ultra-low phase noise of the lasers by having two simultaneous interferometers in an arrangement shown in Fig.\ \ref{conjugate}.  Four frequencies $f_{1}...f_{4}$ are generated from the two phase-locked Ti:Sapphire lasers using an acousto-optic modulator.  Using atoms in different (internal and/or external) states, the individual interferometers are addressed by the beam pairs with the frequencies \{$f_1,f_4$\} and \{$f_2,f_3$\}, respectively.  In many cases, e.g., the photon recoil measurements \cite{hoverm} and the gravity gradient measurements \cite{gravitygradient}, the interesting physics is revealed in the difference of the interferometer phases, which are proportional to the frequency differences $f_1-f_4$ and $f_2-f_3$.  For effective noise reduction, we take the beatnotes of overlapped beams at $f_1-f_2$ and $f_3-f_4$, and phase-lock their sum $(f_1-f_2)+(f_3-f_4)$ by a secondary PLL, which is added to the main PLL by shifting the LO phase.  This, however, is the frequency containing the information, $(f_1-f_4)-(f_2-f_3)$; thus, the laser frequencies are generated such that the differential phase is directly stabilized.

Noise from vibrational variations of the optical path lengths $\delta l_\text{u},\delta l_\text{d}$ from the beat detectors to the atoms for the upper and lower beams (Fig.\ \ref{conjugate}) is effectively cancelled:  In the individual paths, it causes optical phase changes of $2\pi \delta l_\text{u} \frac{f_{1,2}}{c},2\pi\delta l_\text{d} \frac{f_{3,4}}{c}$, which results in phase shifts in the individual atom interferometers proportional to $2\pi(\delta l_\text{u} \frac{f_{1}}{c}-\delta l_\text{d} \frac{f_{4}}{c})$ and $2\pi(\delta l_\text{u} \frac{f_{2}}{c}-\delta l_\text{d} \frac{f_{3}}{c})$.  The difference of the readout phases of the simultaneous interferometers is then proportional to $2\pi(\delta l_\text{u}\frac{f_{1}-f_{2}}{c}-\delta l_\text{d}\frac{f_{4}-f_{3}}{c})$.  Since the wavelengths of the beatnote frequencies $\lambda_{ij}=\frac{c}{f_{i}-f_{j}}$ is large compared to the amplitude of vibrations $\delta l_\text{u},\delta l_\text{d}$, the common-mode noise is strongly suppressed.  We operate such a secondary PLL for the differential phase (Fig.\ \ref{conjugate}), and measure the residual noise in the differential phase using another beat detector at a different point in the optical path. As a result, the noise (dashed line in Fig.\ \ref{noise}) is reduced to $2.5\times10^{-8}$\,rad$^2$, which is servo gain limited.

In summary, we have demonstrated phase-locked \text{1.6\,W} Ti:sapphire lasers with an ultra-low phase noise of \text{-138\,dBc/Hz} at 1\,MHz offset, capable of making instantaneous difference frequency steps of up to 4\,MHz in 200\,ns without cycle slips. We have described the use of this laser system for simultaneous atom interferometers such that the readout of the differential phase of the interferometers can make use of this low noise, rejecting noise from optics vibrations. The will be used to perform atom interferometry for photon-recoil measurements using Cs atoms with part-per-billion accuracy.

We would like to thank Chris Vo for discussions.  \text{H. M.} acknowledges the support from the Alexander von Humboldt Foundation.  This work is sponsored in part by grants from the AFOSR, the NSF and the MURI.




\end{document}